\documentclass[%
 reprint,
 amsmath,amssymb,
 aps,
 pra,
]{revtex4-2}

\usepackage{graphicx}
\usepackage{dcolumn}
\usepackage{bm}
\usepackage{nicefrac}
\usepackage[normalem]{ulem}

\begin{document}

\preprint{APS/123-QED}

\title{Dynamic Stark shift in a Doppler-broadened four-wave mixing}

\author{M. P. M. de Souza}%
 \email{marcopolo@unir.br}
\affiliation{Departamento de F\'{\i}sica, Universidade Federal de Rond\^onia, 76900-726, Ji-Paran\'a, Rond\^onia, Brazil}

\author{A. A. C. de Almeida}

\author{S. S. Vianna}
\affiliation{Departamento de F\'{i}sica, Universidade Federal de Pernambuco, 50670-901, Recife, Pernambuco, Brazil}

\date{\today}

\begin{abstract}
This work presents a theoretical analysis of the Autler-Townes splitting pattern in the four-wave mixing signal generated in a three-level cascade Doppler-broadened system. We employ the density matrix formalism to write the Bloch equations and solve them numerically.
The solutions allow us to compare the response of the upper level population and the generated signal  coherence for homogeneously and non-homogeneously broadened systems.  Our results reveal an AC Stark shift in the nonlinear signal when the frequency of the strong or weak beam is scanned, in contrast to what is observed in the fluorescence.  Furthermore, we present experimental data for the four-wave mixing signal in a hot rubidium vapor for the copropagating configuration of the exciting beams. The behavior of the AC Stark displacement and signal amplitude as a function of laser power indicates a good agreement between the model and the experimental results.
\end{abstract}

\maketitle

\section{Introduction}
\label{introduction}

The dynamic Stark shift, also known as the AC Stark effect, is a critical process in the light-matter interaction subject. It is responsible for the energy level splitting at a transition driven by a near-resonant strong field. One of the consequences in fluorescent spectroscopy is the Mollow triplet \cite{Mollow,Gutierrez}, explained by the split of the ``bare'' atom states into two ``dressed" states whose separation is dependent on the Rabi frequency. In a three-level system, another associated phenomenon is the Autler-Townes (AT) splitting \cite{Autler, Cohen}. In this case, a weak field tuned near one of the transitions, probes the Stark shifted levels stimulated by a strong field laser coupled in the other transition. The result is a double peak in the weak field absorption. This Autler-Townes doublet has been intensively investigated in atomic and molecular spectroscopy. As the splitting amplitude depends on the Rabi frequency of the transitions, the AT splitting has been used to measure transition dipole moments and lifetimes of highly excited states \cite{Piotrowicz, Garcia-Fernandez}. Other applications include quantum memory storage \cite{Saglamyurek} and microwave propagation in transmission lines \cite{Peng}, for instance.

Usually, for a three-level cascade system, a two-beam counterpropagating scheme is used to explore the Doppler-free configuration, especially when the wavelength difference of the two transitions is very small. The first works investigating the AT splitting in a Doppler-broadened medium date from the 1970s \cite{Feneuille, Salomaa, Delsart}. In this case, and for the high-intensity regime, it is challenging to model the system analytically. A more viable alternative is to use numerical methods, as previous papers have done. An example is the work of Ahmed and Lyyra \cite{Ahmed}, which investigated the critical role of Doppler width on the observation of the AT splitting for co- and counter-propagating beams. In particular, they show that the presence of the AT splitting in the fluorescence not only depends on the Rabi frequency but also on the ratio between the wavenumbers of the coupling and probe beams.  Moreover, for copropagating beams, this phenomenon is difficult to be observed. However, as we show in this work, the AT doublet can be distinguished in a four-wave mixing (FWM) using a copropagating setup.

The AT splitting also appears, naturally, in light generation by parametric processes since these typically use high-intensity beams. One of the pioneering works was presented by Boyd \textit{et al}. in 1981 \cite{boyd1981}, in which the signal produced by a FWM process in a two-level system was theoretically studied as a function of the probe beam detuning. The combination of parametric emission and velocity distribution introduces the possibility of quantum interference between oscillating atomic dipoles, such as reported by Zuo and collaborators \cite{Zuo}, a fact that can significantly modify the macroscopic polarization generated by the medium.

In this work, we discuss how the Autler-Townes separation manifests itself in the FWM process at a three-level cascade Doppler-broadened sample. In particular, we are interested in understanding the AT splitting observed in the FWM signal generated when two copropagating beams excite the $5S\rightarrow5P\rightarrow5D$ transitions. One characteristic of this system is that only two transitions are controlled by external beams, whereas the third transition of the FWM process is performed via an amplified spontaneous emission (ASE) process \cite{Akulshin2009}. This system appears as a good candidate in quantum information science and has been successfully used to perform an optical vortex conversion from IR to blue frequencies \cite{Chopinaud}.

In this context, we present theoretical results of the spectrum of the coherent blue light (CBL) and fluorescence generated in rubidium vapor as a function of parameters such as detuning and intensities of the incident beams. Our focus is to understand the contribution of the different atomic group velocities in the AT splitting, whether we scan the frequency of the weak beam or the strong beam. The theoretical analysis allows us to understand the doublet structure observed in the FWM signal. Especially, our experimental results reveal an AC Stark shift in the nonlinear signal independently of which beam we sweep the frequency, in contrast to what is observed in the fluorescence. 

In Sec II, we present a theoretical model using Bloch's equations. We solve these equations numerically with a fourth-order Runge-Kutta method. With the solutions of each term at hand, we analyze the velocity dependence of the FWM output, as well as the fluorescence in Sec III. For this, we investigate both frequency scanning schemes of the weak and strong beams.  Sec. IV is devoted to including the Doppler-broadening and showing its importance to the manifestation of the AT splitting in the experimental signal. The comparison with the experimental data is in Sec. V, in which we also detail the experimental setup. We conclude by summarizing the relevant achievements of this work in Sec. VI.

\section{Atomic system and Bloch equations}
\label{atomic-system}

\begin{figure}[htbp]
	\centering
		\includegraphics[width=0.30\textwidth]{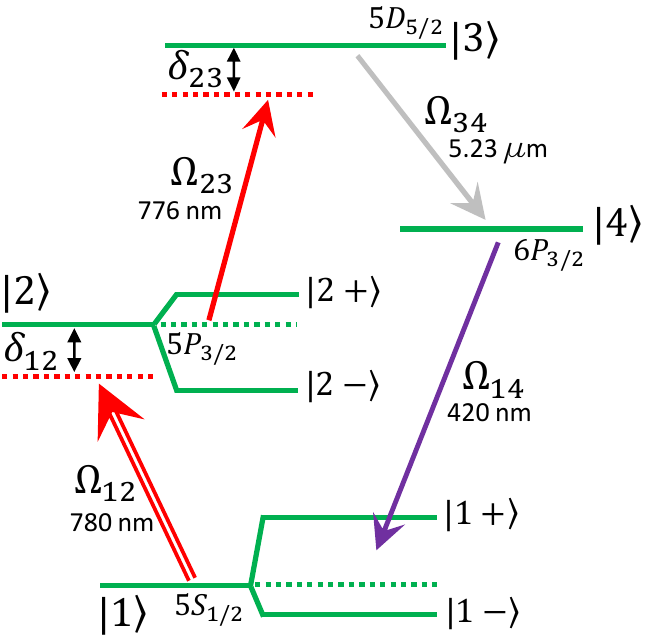}
	\caption{Stark-shifted rubidium level diagram considering a strong field in the $\left|1\right\rangle \rightarrow \left|2\right\rangle$ transition. $\Omega_{ij}$ and $\delta_{ij}$ are the Rabi frequency and the detuning of the fields regarding $\left|i\right\rangle \rightarrow \left|j\right\rangle$ transition, respectively.}
	\label{fig1}
\end{figure}

We model our problem with a four-level system based on the rubidium excitation route $5S_{\nicefrac{1}{2}}\rightarrow5P_{\nicefrac{3}{2}}\rightarrow5D_{\nicefrac{5}{2}}\rightarrow6P_{\nicefrac{3}{2}}$. We label each of these states with a number to simplify the notation, as we show in Fig. \ref{fig1}. Our focus is on the CBL generated near the $\left|4\right\rangle \rightarrow \left|1\right\rangle$ transition (420 nm) when three cw (continuous-wave) input fields with Rabi frequencies $\Omega_{12}$, $\Omega_{23}$, and $\Omega_{34}$ copropagate through the Doppler-broadened rubidium atoms. The field in the $\left|1\right\rangle \rightarrow \left|2\right\rangle$ transition is intense and, therefore, the Stark effect splits the level structure, leading to new resonances associated with the dressed states $\left|1\;\pm\right\rangle$ and $\left|2\;\pm\right\rangle$. These new states have an energy shift given by \cite{boyd2008}

\begin{subequations}
\label{dressed-states}
\begin{eqnarray}
	\omega_{1\pm} &=& -\frac{\delta_{12}}{2} \pm \frac{1}{2}\sqrt{4\Omega_{12}^2 + \delta_{12}^2} ,\\
	\omega_{2\pm} &=& +\frac{\delta_{12}}{2} \pm \frac{1}{2}\sqrt{4\Omega_{12}^2 + \delta_{12}^2}.
\end{eqnarray}
\end{subequations}

We begin the analysis employing the density-matrix formalism with Liouville's equation, 

\begin{equation}
\label{density-matrix}
	\frac{\partial\hat{\rho}}{\partial t} = -\frac{i}{\hslash}\left[\hat{H}, \hat{\rho}\right] + \textrm{decaying terms},
\end{equation}

\noindent where $\hat{\rho}$ is the density matrix operator and $\hat{H}$ is the Hamiltonian of the system including the light-matter interaction in the electric dipole approximation. The matrix representation of the Hamiltonian $\hat{H}$ is

\begin{equation}
\label{hamiltonian}
	\hat{H} = \hslash\left(
	\begin{array}{cccc}
		0 & -\Omega^\prime_{12} & 0 & -\Omega^\prime_{14}\\
		-\Omega^\prime_{21} & \omega_{21} & -\Omega^\prime_{23} & 0 \\
		0 & -\Omega^\prime_{32} & \omega_{31} & -\Omega^\prime_{34} \\
		-\Omega^\prime_{41} & 0 & -\Omega^\prime_{43} & \omega_{41}
	\end{array}
	\right).
\end{equation}

\noindent In this notation, $\omega_{jk} = (E_j - E_k)/\hslash$ are the resonance frequencies of the $\left|k\right\rangle \rightarrow \left|j\right\rangle$ transitions and $\Omega^\prime_{jk} \equiv \Omega_{jk}(e^{i\omega_{kj}t} + e^{-i\omega_{kj}t})$. For a group of atoms with velocity component $v$ in the direction of the propagation of the lasers, we write the Bloch equations in the rotating wave approximation as

\begin{subequations}
\label{bloch}
\begin{eqnarray}
\dot{\rho}_{11} &=& -i\Omega_{12}\sigma_{12} + c.c. - i\Omega_{14}\sigma_{14} + c.c + \nonumber
\\
           &+& \Gamma_{22}\rho_{22} + \Gamma_{41}\rho_{44},
\\
\dot{\rho}_{22} &=& i\Omega_{12}\sigma_{12} + c.c. - i\Omega_{23}\sigma_{23} + c.c. - \nonumber
\\
           &-& \Gamma_{22}\rho_{22} + \Gamma_{32}\rho_{33},
\\
\dot{\rho}_{33} &=& i\Omega_{23}\sigma_{23} + c.c. - i\Omega_{34}\sigma_{34} + c.c. - \nonumber
\\
           &-& (\Gamma_{32}+\Gamma_{34})\rho_{33},
\\
\dot{\rho}_{44} &=& -i\Omega_{34}\sigma_{23} + c.c. + i\Omega_{14}\sigma_{14} + c.c. + \nonumber
\\
           &+& \Gamma_{34}\rho_{33} - \Gamma_{41}\rho_{44},
\\
\dot{\sigma}_{12} &=&  \left[i(\delta_{12} - k_{12}v) - \gamma_{12} \right]\sigma_{12} + i\Omega_{14}\sigma_{42} - \nonumber
\\
					  &-& i\Omega_{23}\sigma_{13} + i\Omega_{12}(\rho_{22} - \rho_{11}),
\\
\dot{\sigma}_{23} &=&  \left[i(\delta_{23} - k_{23}v) - \gamma_{23} \right]\sigma_{23} + i\Omega_{12}\sigma_{13} - \nonumber
\\
            &-& i\Omega_{43}\sigma_{24} + i\Omega_{23}(\rho_{33} - \rho_{22}),
\\
\dot{\sigma}_{14} &=&  \left[i(\delta_{14} - k_{14}v) - \gamma_{14} \right]\sigma_{14} + i\Omega_{12}\sigma_{24} - \nonumber
\\
            &-& i\Omega_{43}\sigma_{13} + i\Omega_{14}(\rho_{44} - \rho_{11}),
\\
\dot{\sigma}_{43} &=&  \left[i(\delta_{43} - k_{43}v)  - \gamma_{43} \right]\sigma_{43} + i\Omega_{14}\sigma_{13} - \nonumber
\\
            &-& i\Omega_{23}\sigma_{42} + i\Omega_{43}(\rho_{33}-\rho_{44}),
\\
\dot{\sigma}_{13} &=&  \left[i(\delta_{12}+\delta_{23} - (k_{12} + k_{23})v) - \gamma_{13} \right]\sigma_{13} + \nonumber
\\
             &+& i\Omega_{12}\sigma_{23} + i\Omega_{14}\sigma_{43} - i\Omega_{23}\sigma_{12}-i\Omega_{43}\sigma_{14},
\\
\dot{\sigma}_{24} &=&  \left[i(\delta_{14}-\delta_{12} - (k_{14}- k_{12})v) - \gamma_{24} \right]\sigma_{24} + \nonumber
\\
             &+& i\Omega_{12}\sigma_{14} + i\Omega_{23}\sigma_{34} - i\Omega_{14}\sigma_{21} - i\Omega_{43}\sigma_{23}.
\end{eqnarray}
\end{subequations}

\noindent where $\gamma_{jk}$ represent the relaxation rates of the coherences, $\Gamma_{jk}$ is the population spontaneous relaxation rate from a $\left|k\right\rangle$ state to a $\left|j\right\rangle$ state, $\delta_{jk}$ and $k_{jk}$ are the detuning and wavenumber of the field associated to the Rabi frequency $\Omega_{jk}$ and $\sigma_{jk} \equiv \rho_{jk}e^{-i\omega_{jk} t}$ is the slow envelope of the coherence $\rho_{jk}$.

Out of all elements of the density matrix, we are interested in the coherence $\sigma_{14}$ and the population of the state $\left|3\right\rangle$, $\rho_{33}$. The square modulus of $\sigma_{14}$ renders the  FWM signal while the upper state population represents the fluorescence emitted by the atoms. To compare the results with the experimental data, one must take into account the contribution of all velocity groups of the hot vapor:

\begin{subequations}
\begin{eqnarray}
\label{rho-sigma}
	\bar{\rho}_{33} &=& \int_{-\infty}^{\infty}{\rho_{33}(v)f(v)dv} , \\
	\bar{\sigma}_{14}, &=& \int_{-\infty}^{\infty}{\sigma_{14}(v)f(v)dv},
\end{eqnarray}
\end{subequations}

\noindent where $f(v)$ is the Maxwell-Boltzmann velocity distribution.

The results presented in the following sections were obtained by solving the Bloch equations (Eqs. \ref{bloch}) numerically using the fourth-order Runge-Kutta method from time $t = 0$ to $t = 2$ $\mu s$. The integration time is approximately the average transit time of the atoms through laser beams with a 0.2 mm diameter (similar to the experimental configuration described in Sec. V). The initial conditions are $\rho_{11}(0) = 1$ and $\rho_{22}(0) = \rho_{33}(0) = \rho_{44}(0) = \sigma_{ij}(0) = 0$. Since this is a very demanding calculation, we use three graphics processing units (\textit{Nvidia RTX 2070 Super}) to solve the differential equations for all velocity groups simultaneously. The numerical parameters we use in all computations of this work are in table \ref{tab:table1}. To simulate the amplified spontaneous emission that occurs in the $5D\rightarrow 6P$ transition \cite{Akulshin2009}, we chose a small Rabi frequency value for the 5.23 $\mu$m field ($\Omega_{34}$ = 1 rad/s) to act as a seed in the FWM process. Furthermore, we fix the frequency of this field on resonance ($\delta_{34} = 0$) and, consequently, $\delta_{14} = \delta_{12} + \delta_{23} - \delta_{34}$ due to the energy conservation.

\section{Velocity dependence of \textbf{the population $\rho_{33}$ and} the FWM signal}
\label{velocity}

It is interesting that, before adding the contribution of all atoms, introducing the Doppler-broadening, we investigate the dependence of the fluorescence and the FWM signal with different individual atomic velocity groups. We consider two frequency scanning scenarios, sweeping either the frequency of the weak beam ($\Omega_{23}$) or the frequency of the strong beam ($\Omega_{12}$).

\begin{table}
\caption{\label{tab:table1} Numerical parameters for the theoretical model.}
\begin{ruledtabular}
\begin{tabular}{lr}
Decay rates (MHz)\\
\hline
$\Gamma_{22}$ & $2\pi\times 6.06$\footnotemark[1]\\
$\Gamma_{33}$ & $2\pi\times 0.66$\footnotemark[1]\\
$\Gamma_{44}$ & $2\pi\times 1.3$\footnotemark[1]\\
$\Gamma_{32}$ & $0.65\Gamma_{33}$\\
$\Gamma_{34}$ & $0.35\Gamma_{33}$\\
$\gamma_{12}$ & $\Gamma_{22}/2$\\
$\gamma_{23}$ & $(\Gamma_{33}+\Gamma_{22})/2$\\
$\gamma_{34}$ & $(\Gamma_{33}+\Gamma_{44})/2$\\
$\gamma_{14}$ & $\Gamma_{44}/2$\\
$\gamma_{13}$ & $\Gamma_{33}/2$\\
$\gamma_{24}$ & $(\Gamma_{44}+\Gamma_{22})/2$\\
\hline
Wavelenghts (nm)\\
\hline
$\lambda_{12}$ & $780$\\
$\lambda_{23}$ & $776$\\
$\lambda_{34}$ & $5300$\\
$\lambda_{14}$ & $420$\\
\end{tabular}
\end{ruledtabular}
\footnotetext[1]{Data from Ref. \cite{steck}.}
\end{table}

\subsection{Weak beam frequency sweeping}
\begin{figure}[htbp]
	\centering
		\includegraphics[width=0.45\textwidth]{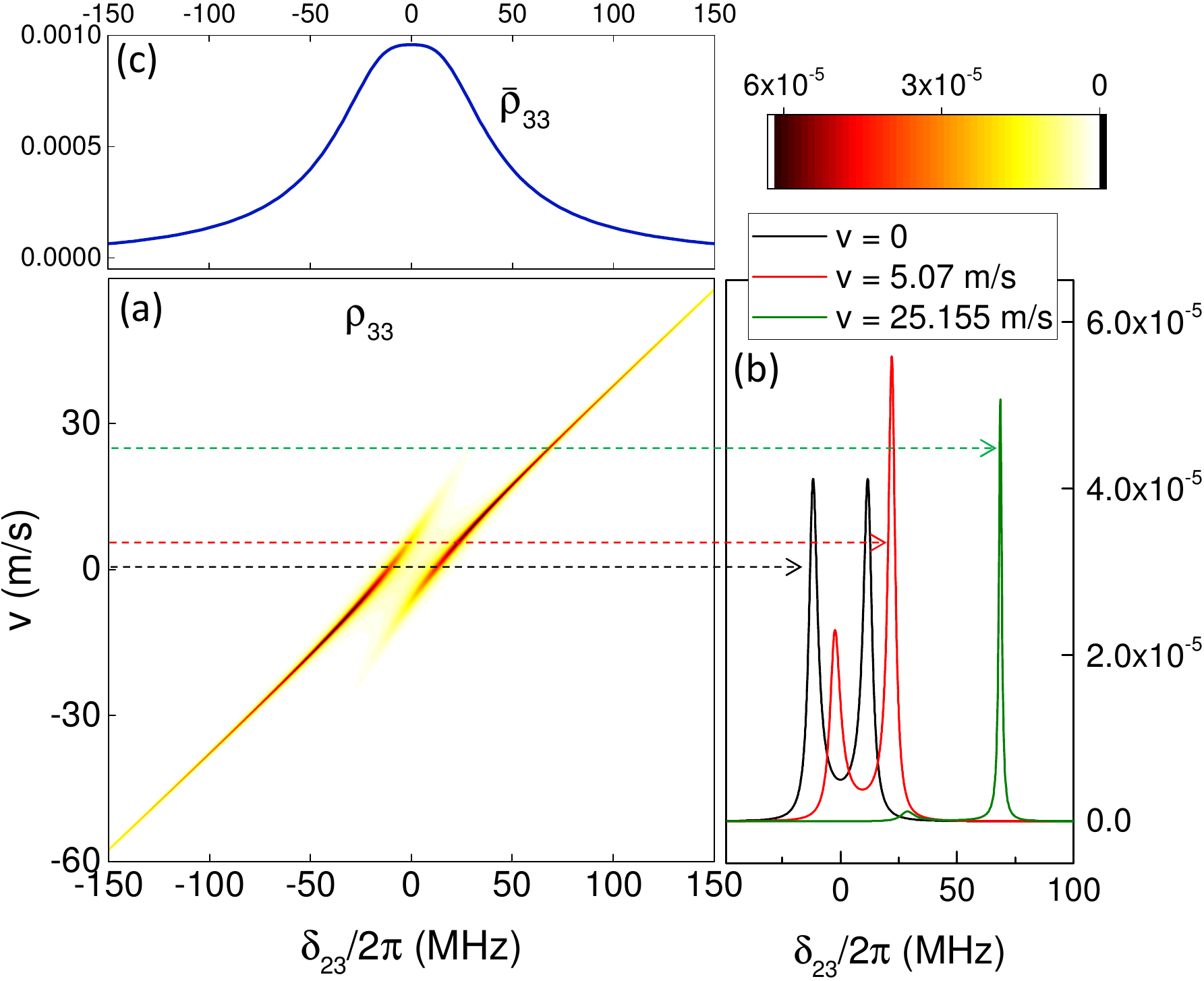}
	\caption{(a) $\rho_{33}$ as a function of $\delta_{23}$ and $v$, considering $\Omega_{12}/(2\pi) = 12$ MHz, $\Omega_{23}/(2\pi) = 0.6$ MHz and $\delta_{12} = 0$. (b) $\rho_{33}$ for three velocity groups. (c) $\rho_{33}$ integrated over the Maxwell-Boltzmann distribution.}
	\label{fig2}
\end{figure}

The first case we analyze is when the frequency of weak beam $\Omega_{23}$ is scanning while the strong beam is on resonance with the $\left|1\right\rangle \rightarrow \left|2\right\rangle$ transition ($\delta_{12} = 0$) for the zero velocity group.
In Fig. \ref{fig2}(a), we present the population of the upper state $\rho_{33}$ as a function of $\delta_{23}$ for many velocity groups. 

The behavior of the population $\rho_{33}$ for three specific velocity groups is highlighted in Fig. 2(b), where it is possible to see the doublet structure due to the Autler-Townes effect \cite{Autler}. Considering $k_{12} \approx k_{23} \equiv k$, the peaks of the doublets shown in Fig. \ref{fig2}(b)  arise when the two-photon condition is satisfied. Therefore, we solve the equation $\delta_{23} - kv + \omega_{2\pm} = 0$ and find that (\cite{Feneuille})

\begin{equation}
\label{delta23}
	\delta_{23} = \frac{3kv}{2} \pm \frac{1}{2}\sqrt{4\Omega_{12}^2 + k^2v^2}.
\end{equation}

\noindent Whenever $v \neq 0$, the doublet is asymmetric as the red curve in Fig. \ref{fig2}(b) indicates. Furthermore, the nearest side-band from the resonance is always smaller \cite{Gray}. Although specific velocity groups might have a level splitting, the Doppler-broadening hides the AT effect, as it is possible to observe in Fig. \ref{fig2}(c). This result is also in agreement with the discussion in Ref. \cite{Feneuille}: one should observe a larger power broadening without any splitting in the fluorescence signal. Consequently, it is unlikely to observe the splitting in experiments that detect only fluorescence. However, as we will show, the FWM can reveal this effect even after the Doppler integration.

\begin{figure}[htbp]
	\centering
		\includegraphics[width=0.45\textwidth]{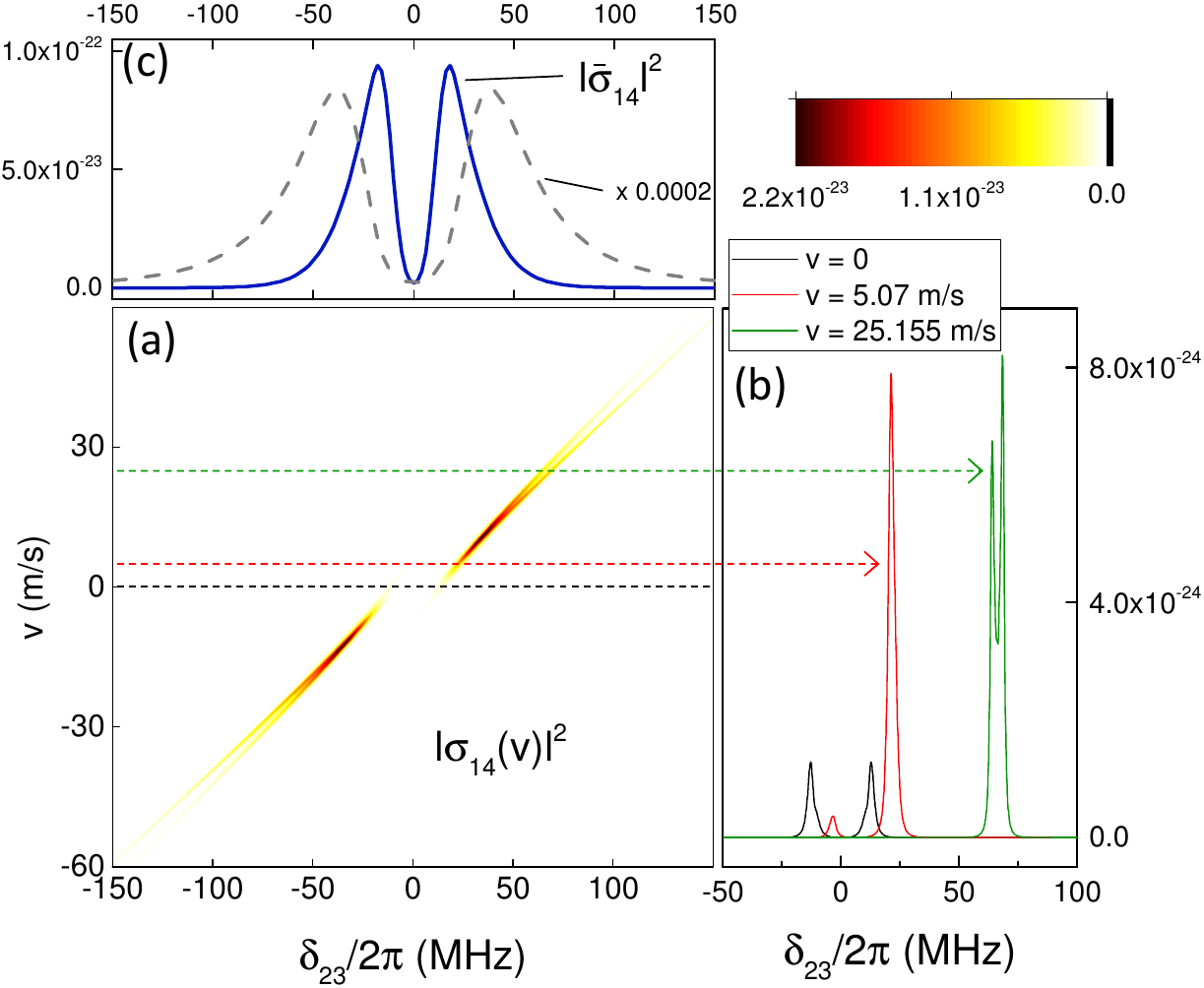}
	\caption{(a) $\lvert\sigma_{14}(v)\rvert^2$ as a function of $\delta_{23}$, for many velocity groups, considering $\Omega_{12}/2\pi = 12$ MHz, $\Omega_{23}/2\pi = 0.6$ MHz  and $\delta_{12} = 0$. (b) $\lvert\sigma_{14}(v)\rvert^2$ for three velocity groups. (c) Blue: $\sigma_{14}$ integrated over the Maxwell-Boltzmann distribution and then squared; Dashed line: $\lvert\sigma_{14}\rvert^2$ integrated over the Maxwell-Boltzmann distribution}
	\label{fig3}
\end{figure}

In Fig. \ref{fig3}, we present a similar analysis for the coherence $\sigma_{14}$, that is, we first look at the response of each velocity group and then get the total response of the Doppler-broadening medium.  In this case, as we are interested in the FWM signal, we look not at the coherence but its squared modulus. The colormap of Fig. \ref{fig3}(a) presents a doublet structure, as did Fig. \ref{fig2}(a), but with a different shape. To better understand this curve, we can look at a few specific velocity groups in Fig. \ref{fig3}(b). Each of these curves has its peak positions dictated by Eq. (\ref{delta23}), connecting its origin to the FWM process.

It is important to note that we present the square modulus of $\sigma_{14}$ to each velocity group in Fig. \ref{fig3}(a) and (b). However, to plot the actual FWM signal, one must first integrate the coherence $\sigma_{14}$ with the Maxwell-Boltzmann distribution to, only then, take its squared modulus, as shown in the blue curve of Fig. \ref{fig3}(c). We perform the calculations in this particular order to avoid neglecting the phases between each velocity group. As discussed in Ref \cite{Zuo}, the macroscopic field polarization from different ensembles of atoms within the atomic velocity groups can interfere, causing a significant modification of the FWM spectra. To support this argument, we present the dashed curve of Fig. \ref{fig3}(c), in which first we take the squared modulus of the coherence and then integrate it. Notice that, even though the splitting is present, the frequency distance between the peaks is higher than in the blue curve. 

Comparing Figs. \ref{fig2}(b) and \ref{fig3}(b), one can see that the behavior of the coherence $\sigma_{14}$ is much different from the population $\rho_{33}$ when it comes to the nonzero velocity groups (see the green and red curves). Once the velocity increases, there is a dramatic difference in the coherence response near resonance. The AC Stark effect takes place and splits the level structure for each velocity group, as in the red curve of Fig. \ref{fig3}(b). However, one of the peaks, the one near resonance, is very small. Once all the groups add up in the Doppler integration, they contribute to the FWM signal in the blue curve of Fig. \ref{fig3}(c) with a doublet-like structure. The same argument applies to the population $\rho_{33}$. However, in this case, the difference between the peaks amplitude is not as dramatic as with the coherence, so once the groups add up, the splitting is gone.

\subsection{Strong beam sweeping}

Here we analyze an unusual case, where the presence of Stark shift is investigated as a function of the strong field detuning, $\delta_{12}$. Using again Eq. (\ref{dressed-states}b) and the resonance of two photons with the frequency of the weak field fixed at the resonance ($\delta_{23} = 0$), we obtain the following equation for the peak position, considering $v \neq 0$:

\begin{equation}
\label{delta-12}
	\delta_{12} = \frac{2k^2v^2 - \Omega_{12}^2}{kv}.
\end{equation}

\begin{figure}[htbp]
	\centering
		\includegraphics[width=0.45\textwidth]{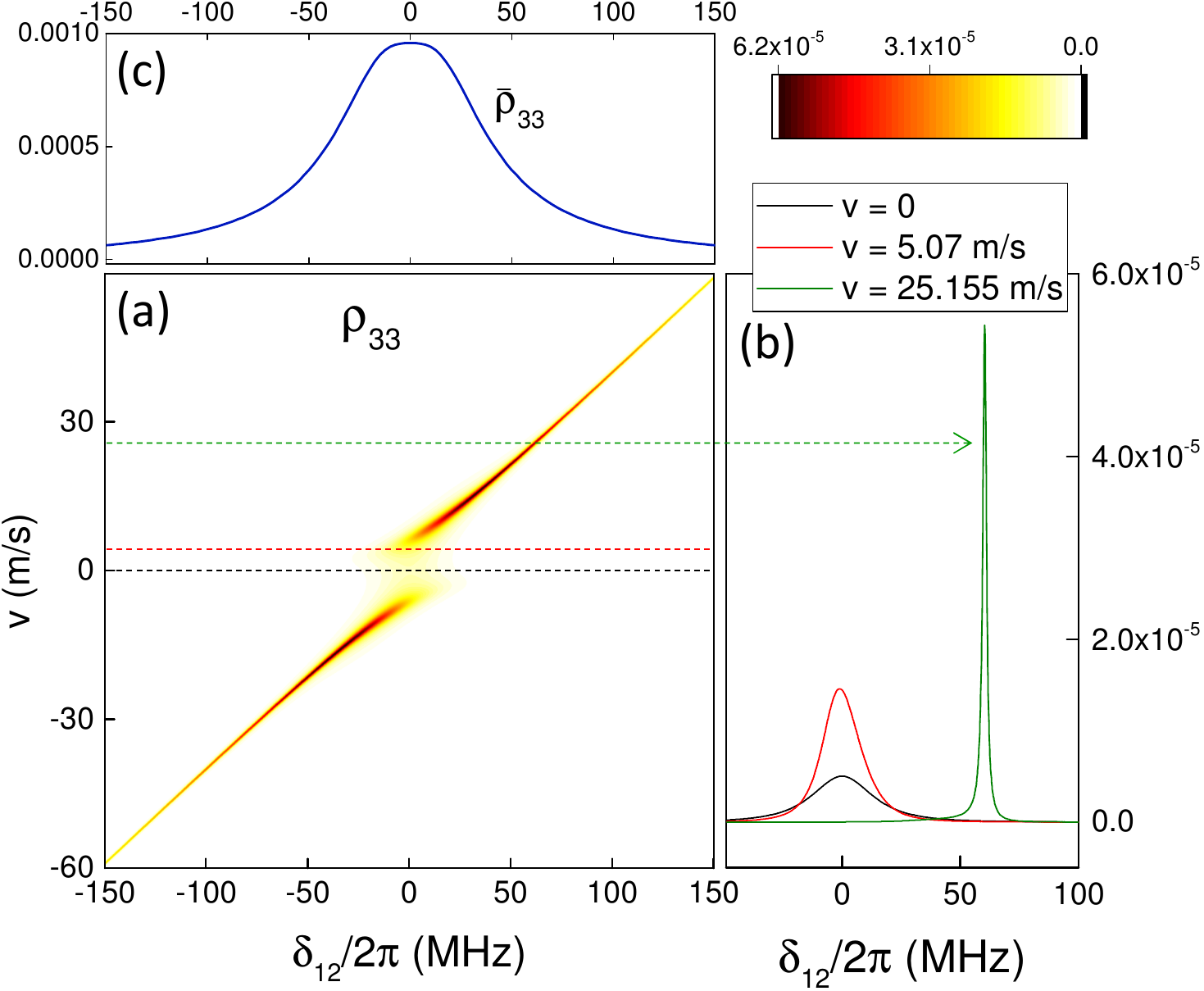}
	\caption{(a) $\rho_{33}$ as a function of $\delta_{12}$ and $v$, considering $\Omega_{12}/(2\pi) = 12$ MHz, $\Omega_{23}/(2\pi) = 0.6$ MHz and $\delta_{23} = 0$. (b) $\rho_{33}$ for three velocity groups. (c) $\rho_{33}$ integrated over the Maxwell-Boltzmann distribution.}
	\label{fig4}
\end{figure}

We present the results of the upper population $\rho_{33}$ in Figs. \ref{fig4}(a) and (b), as a function of $\delta_{12}$ considering $\Omega_{12} = 2\pi\times 12$ MHz, $\Omega_{23} = 2\pi \times 0.6$ MHz and $\delta_{23} = 0$. It is interesting to notice that the two-photon resonance condition for $v = 0$ is never satisfied, and it is necessary a far detuned $\Omega_{12}$ field if $v$ is very small ($v < 3$ m/s).  The $\rho_{33}$ population represented by the black and red curves in Fig. \ref{fig4}(b) comes from a non-resonant situation because the $\Omega_{23}$ field does not resonate with the Stark shifted sidebands $\omega_{2\pm}$. The green curve, on the other hand, comes from an exact two-photon resonance, which implies a higher thin peak. As Eq. (\ref{delta-12}) has a single solution, there is not a doublet-like structure. When the Doppler integration is taken into account, the contribution of the non-resonant two-photon transition near $\delta_{12} = 0$ for many velocity groups is enough to result in a large value of $\bar{\rho}_{33}$ near resonance, which hides the Stark shift effect, as can be seen in Fig. \ref{fig4}(c).

The absence of the AT splitting in the upper population of a Doppler medium is due to the copropagating setup. This means that an experimental fluorescence measurement could not show the characteristic two peaks. This is a known fact: the observation of the AT splitting depends on the relation between the wavevectors and the intensities of the incident beams, as pointed by Feneuille and Schweighofer \cite{Feneuille}. In the typical counterpropagating setup, one can observe the AT splitting no matter which beam is sweeping, the strong or the weak \cite{Martinez}. On the other hand, if the atoms are at rest, the behavior is the same for co- and counter-propagating beams.

\begin{figure}[htbp]
	\centering
		\includegraphics[width=0.45\textwidth]{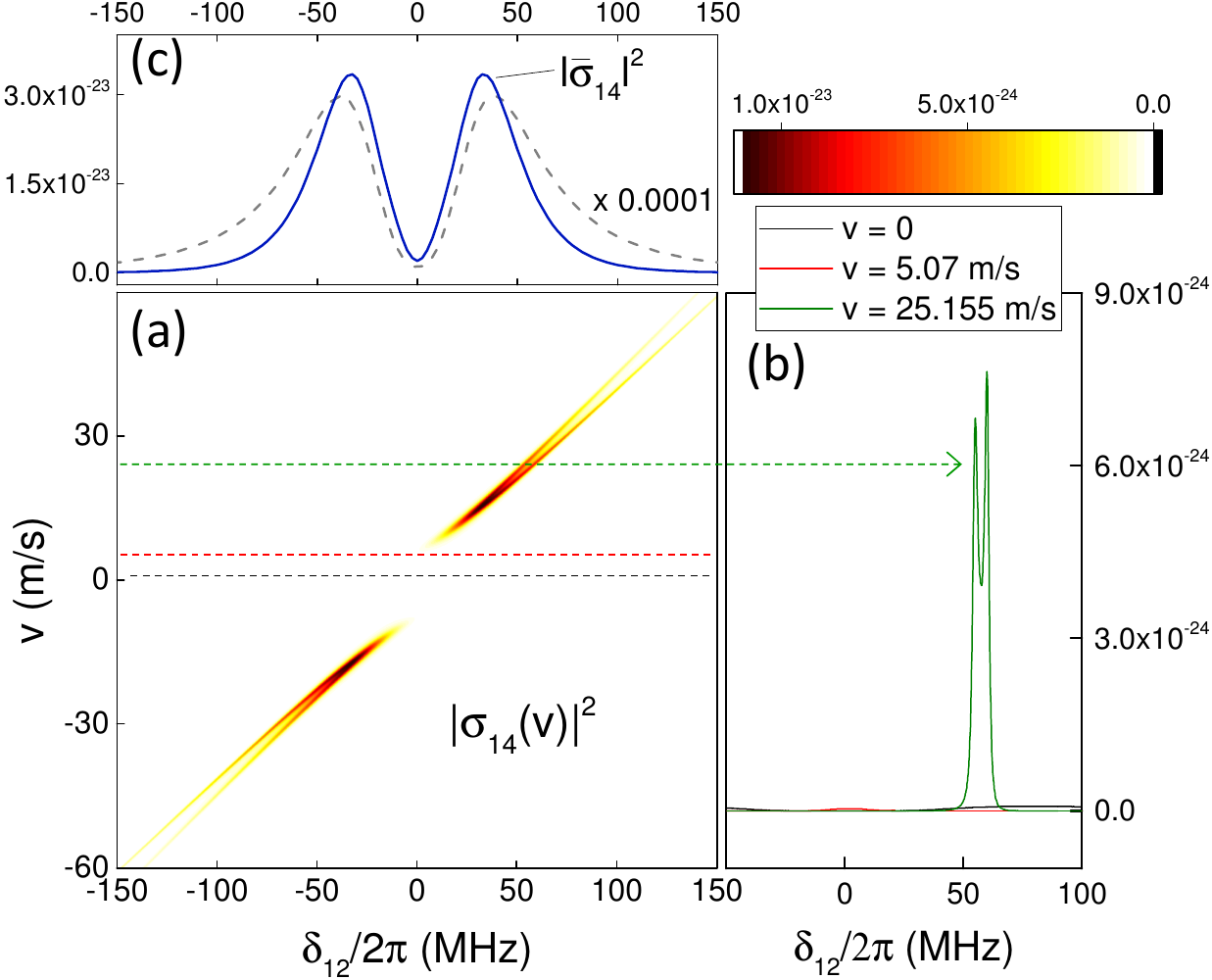}
	\caption{(a) $\lvert\sigma_{14}(v)\rvert^2$ as a function of $\delta_{12}$, for many velocity groups, considering $\Omega_{12}/2\pi = 12$ MHz, $\Omega_{23}/2\pi = 0.6$ MHz  and $\delta_{23} = 0$. (b) $\lvert\sigma_{14}(v)\rvert^2$ for three velocity groups. (c) Blue: $\sigma_{14}$ integrated over the Maxwell-Boltzmann distribution and then squared; Dashed line: $\lvert\sigma_{14}\rvert^2$ integrated over the Maxwell-Boltzmann distribution}
	\label{fig5}
\end{figure}

As in the last subsection, we present the results for the coherence $\sigma_{14}$ in Fig. \ref{fig5}, using the same parameters of Fig. \ref{fig4}. Again, we look not at the coherence but its squared modulus. In comparison with $\rho_{33}$, $\sigma_{14}$ is even more dependent on the velocities of the atoms, leading to an insignificant value near $v = 0$, as Figs. \ref{fig5}(a) and (b) show. Thus, it follows that once the Doppler integration is performed (in the order we discussed previously), the doublet structure will prevail. Then, the hole in the $\left|\bar{\sigma}_{14}\right|^2$, the blue curve of Fig. \ref{fig5}(c), is a consequence of the Stark shift together with a non-satisfied two-photon transition for low velocities.

Notice that in this frequency sweeping regime, the phase between velocity groups is not as relevant to the distance between the two peaks as in the last subsection. The dashed line of Fig. \ref{fig5}(c), obtained first taking the squared modulus of the coherence and then integrating it, is slightly wider than the blue curve. Nevertheless, the following results have the same calculation procedure of the blue curves of Figs. \ref{fig3}(c) and \ref{fig5}(c). 

It is noteworthy that while for copropagating beams, the results for fluorescence and FWM are different whether it is the strong or the weak beam that is scanning, for a configuration of counterpropagating beams, the AT splitting can be observed both in fluorescence and in FWM, no matter which beam is sweeping \cite{Zhang}.

It seems that different physical mechanisms play a role in the presence of the doublet structure in the FWM signal (and the lack of it in the fluorescence) once the Doppler integration is performed for the two frequency sweeping mechanisms. We explore these mechanisms more deeply and highlight their differences in the following section.

\section{The FWM signal after the Doppler integration}

\begin{figure}[htbp]
	\centering
		\includegraphics[width=0.4\textwidth]{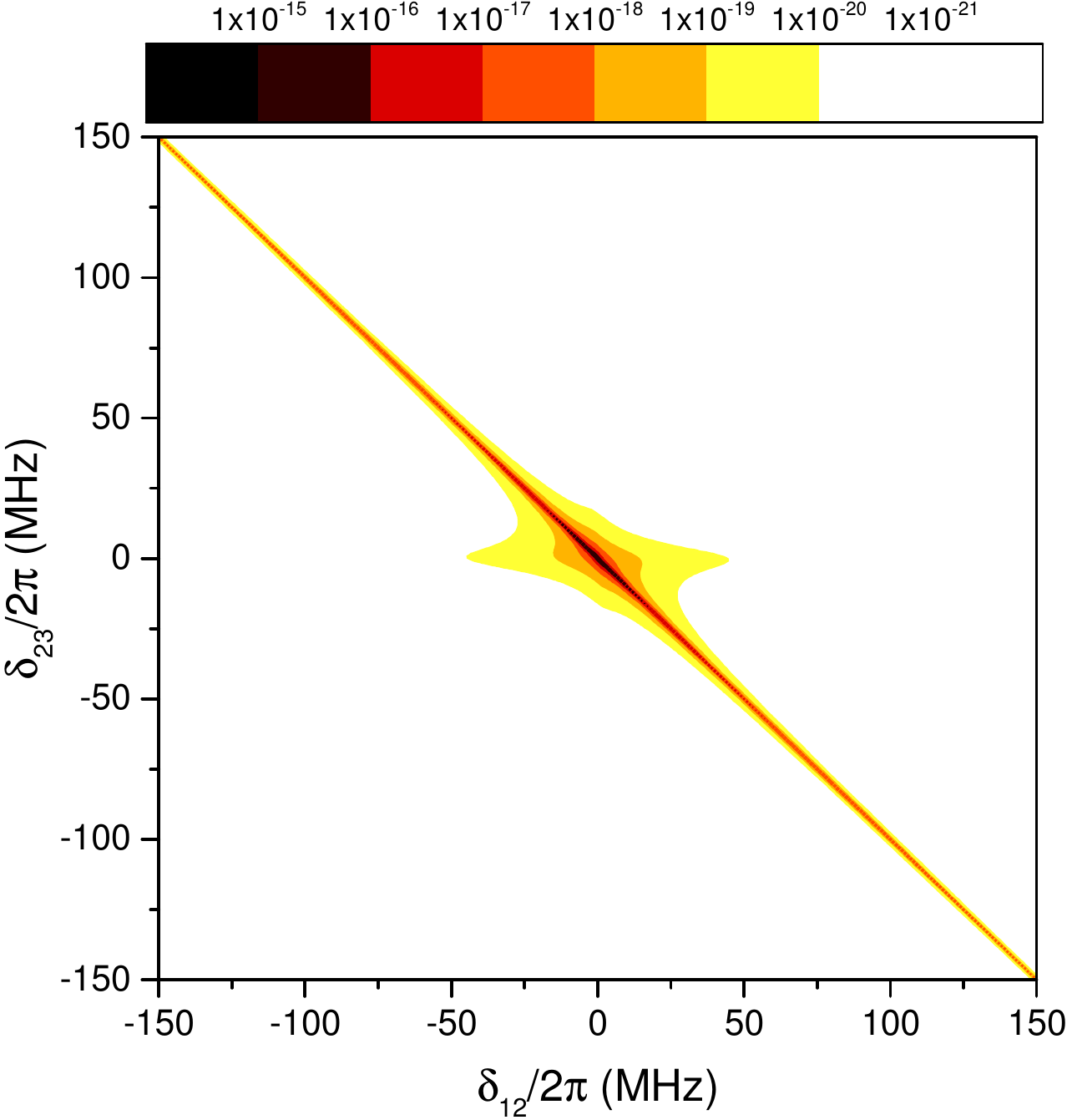}
	\caption{Coherence $\left|\sigma_{14}\right|^2$ as a function of $\delta_{12}$ and $\delta_{23}$, for a group of atoms with $v = 0$ in a weak field regime: $\Omega_{12} = \Omega_{23} = 2\pi\times 0.6$ MHz. Logarithmic scale.}
	\label{fig6}
\end{figure}

For a group of atoms with $v = 0$, there are infinite combinations of laser frequencies that satisfy the two-photon resonance condition in the transition  $\left|1\right\rangle \rightarrow \left|3\right\rangle$. These combinations are given by the equation $\delta_{12} + \delta_{23} = 0$. However, as it is possible to observe in Fig. \ref{fig6}, with $\left|\sigma_{14}\right|^2$ as a function of the laser detunings $\delta_{12}$ and $\delta_{23}$, this coherence is dominated by the two-photon transition when one-photon resonances are also present ($\delta_{12} = \delta_{23} = 0$). Notice that the colormap of Fig. \ref{fig6} is on a logarithmic scale.

\begin{figure}[htbp]
	\centering
		\includegraphics[width=0.45\textwidth]{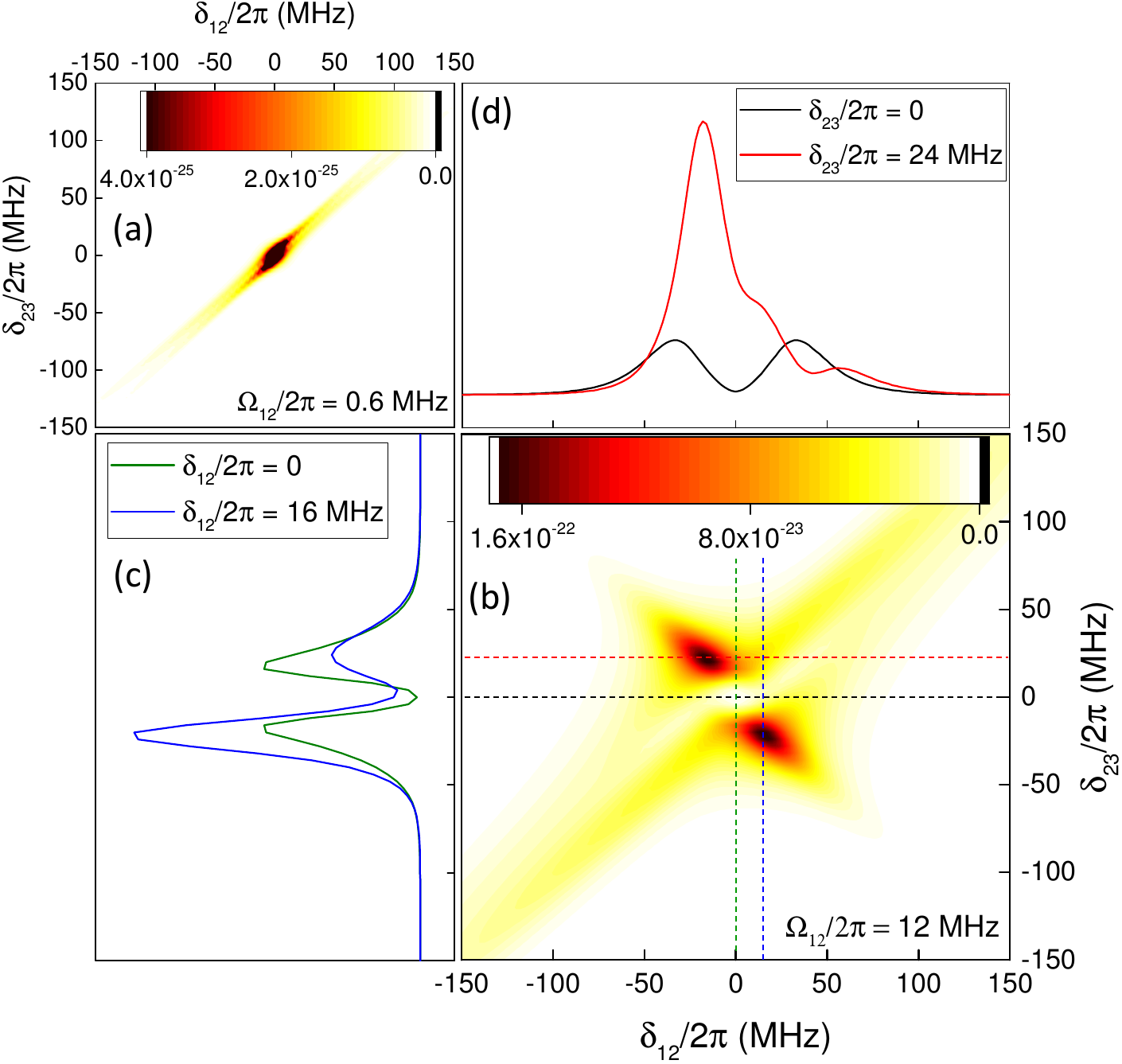}
	\caption{$\left|\bar{\sigma}_{14}\right|^2$ as a function of $\delta_{12}$ and $\delta_{23}$, for (a) $\Omega_{12}/2\pi = 0.6$ MHz and (b) $\Omega_{12}/2\pi = 12$ MHz. (c) and (d): $\left|\bar{\sigma}_{14}\right|^2$ for $\delta_{12}$ or $\delta_{23}$ constant, as indicated by the dashed curves in (b). The green curve in (c) and the black curve in (d) are the same presented in figures 3(c) and 5(c), respectively.}
	\label{fig7}
\end{figure}

The integration over the Maxwell-Boltzmann distribution of velocities introduces even more possibilities. In Fig. \ref{fig7}(a), we present $\left|\bar{\sigma}_{14}\right|^2$ for the same intensities of Fig. \ref{fig6},  but with the integration with the Maxwell-Boltzmann distribution as discussed previously. The configuration $\delta_{12} = \delta_{23} = 0$ dominates again, but for a copropagating beams configuration, $\left|\bar{\sigma}_{14}\right|^2$ is much higher when $\delta_{12} = \delta_{23}$ in comparison with $\delta_{12} = -\delta_{23}$. There is always a group of atoms, with $v \approx \delta_{23}/k_{23}$, which are simultaneously in one- and two-photon resonance when the condition $\delta_{12} = \delta_{23}$ is selected, since $k_{23} \approx k_{12}$ for the rubidium structure levels.

If one increases the Rabi frequency of the field in the lower transition ($\left|1\right\rangle \rightarrow \left|2\right\rangle$) there is a significant impact in the FWM due to the Stark shift, as Fig. \ref{fig7}(b) shows. We plot $\left|\bar{\sigma}_{14}\right|^2$ as a function of $\delta_{12}$ and $\delta_{23}$ when the lower transition is driven by a strong field ($\Omega_{12}/2\pi = 12$ MHz) while the weaker field ($\Omega_{23}/2\pi = 0.6$ MHz) stimulates the upper transition $\left|2\right\rangle \rightarrow \left|3\right\rangle$. It is possible to observe that the single peak in Fig. 7(a) splits into two peaks located over the line $\delta_{12} = -\delta_{23}$. The consequence is the presence of a doublet structure in the typical configuration (weak beam varying its frequency with the strong beam on resonance) and in the opposite situation (strong beam scanning with the weak beam fixed on resonance), as Figs.  \ref{fig7}(c) and (d) demonstrate. Notice that, if the fixed frequency laser is resonant with a different velocity group of atoms, the signal becomes asymmetric.

It is important to remember the results of the previous section. The AT splitting appears for each group of atoms in both scanning configurations in the coherence $\left|\sigma_{14}(v)\right|^2$, but the final FWM spectra can only be achieved after the proper Doppler integration. Therefore, the splitting we observe in the FWM is not merely due to the AC Stark effect but is also connected to the Doppler integration as the velocity groups give different contributions to the final FWM output (see Figs. 3(c) and 5(c)).

\begin{figure}[htbp]
	\centering
		\includegraphics[width=0.45\textwidth]{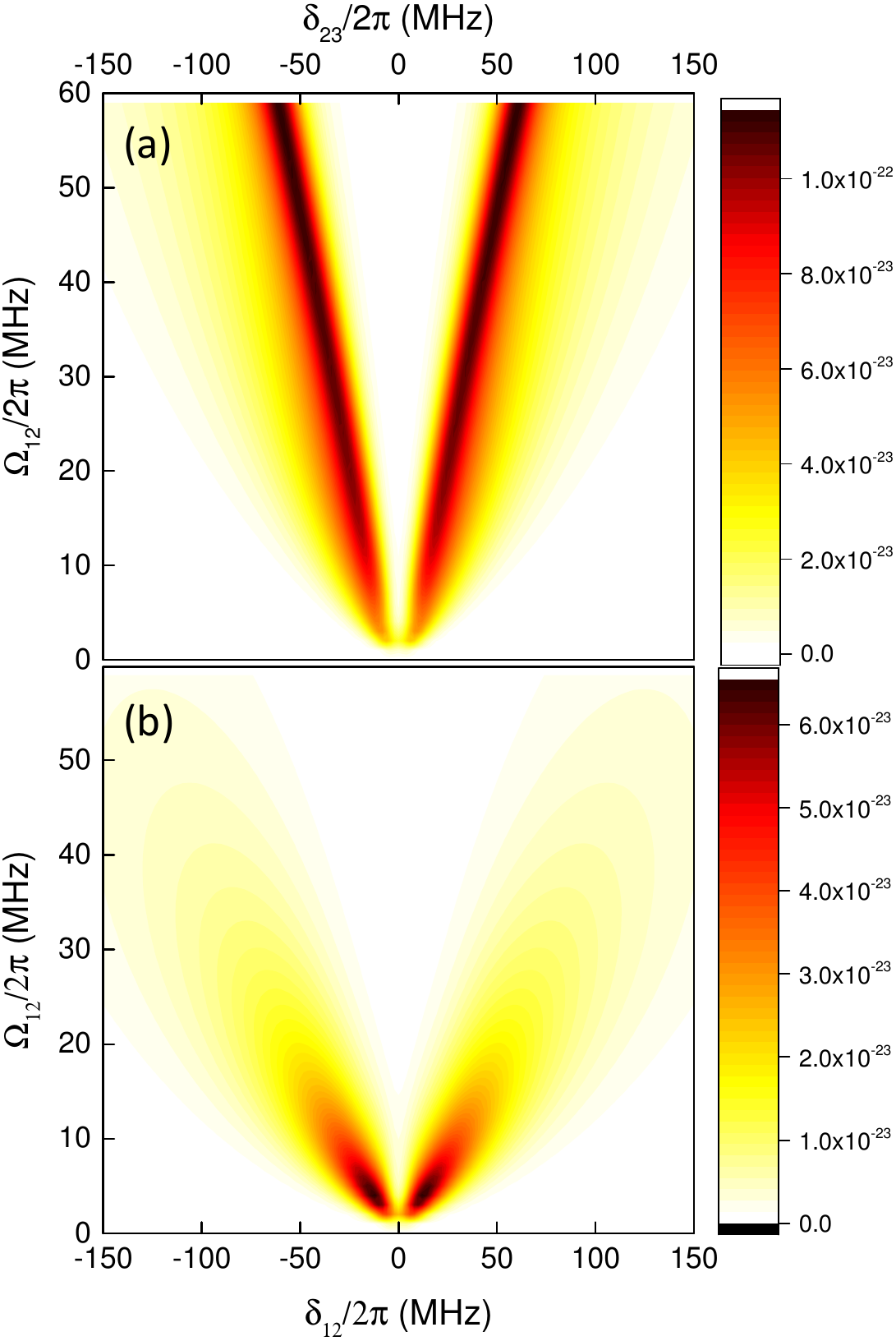}
	\caption{ $\left|\bar{\sigma}_{14}\right|^2$ as a function of $\Omega_{12}$ and (a) $\delta_{23}$ or (b) $\delta_{12}$.}
	\label{fig8}
\end{figure}

The results of Figs. \ref{fig8}(a) and (b) explore more deeply the relation between the FWM signal ($\left|\bar{\sigma}_{14}\right|^2$) and the intensity of the beam at the lower transition, $\Omega_{12}$, for the two frequency scanning setups. Notice that, in both cases, the splitting increases linearly but with different angular coefficients. For the weak beam sweeping scenario, the splitting is $\approx 1.9\Omega_{12}$, very close to the Autler-Townes separation for a single group of atoms with $v = 0$ ($2\Omega_{12}$). On the other hand, if the strong beam frequency varies, the splitting is much higher, $\approx 4.2\Omega_{12}$. We can interpret this greater separation by looking again at the results in Fig. \ref{fig5}(b) for this scanning configuration. It is clear that the contribution to $\left|\bar{\sigma}_{14}\right|^2$ from one of the AT doublet peaks, which is closer to resonance, is negligible. So, this greater separation corresponds to the distance between the most distant peaks of the resonance due to two AT doublets of different velocity groups. In the following section we discuss that the experimental results could not show much difference between the two frequency scanning configurations.  However, a small change in the model parameters can improve the agreement.

Concerning the peak amplitude of the doublet, the behavior is also different: it saturates when the weak beam is sweeping its frequency, while it has a maximum and then decreases when the strong beam is sweeping. There are distinct physical mechanisms in play in these two frequency scanning configurations. In the typical experiment, with the weak beam scanning, the strong field in the lower transition splits the intermediate level into two due to the AC Stark effect. Then, the weak beam in the upper transition probes these split levels. Since we consider a closed system, there will always be atoms that can satisfy the two-photon resonance and induce the FWM process. For the other frequency scanning setup, the weak beam has a fixed frequency, so eventually, the strong beam will lead to a splitting so large that the two-photon resonance can no longer happen, and therefore decreasing the signal.

\section{Comparison with experimental results}
\label{comparison-results}

In the previous sections, we discussed the theoretical model for several configurations. We focus now on experimental results of the FWM process that cover some of these theoretical situations. Particularly, we use an experimental setup with two beams of different lasers copropagating through an Rb vapor at $\approx 80 ^o$C. The first beam comes from a homemade cw diode laser at 780 nm, with stabilized temperature and 30 mW of maximum output power. This laser excites the 5S$_{1/2}, F=3 \rightarrow$ 5P$_{3/2}, F=4$ transition ($\left|1\right\rangle \rightarrow \left|2\right\rangle$) of $^{85}$Rb. The detuning around this transition can reach a maximum of 10 GHz due to the current control of the diode laser. The second beam comes from a mode-locked Ti:sapphire laser (\textit{BR Labs Ltda}) with a repetition rate of $f_R \approx 1$ GHz, maximum output average power of 500 mW, and a 20 nm bandwidth centered near 776 nm. We tune one optical mode of the frequency comb in the Doppler-broadened 5P$_{3/2},F=4 \rightarrow$ 5D$_{5/2},F=5$ transition ($\left|2\right\rangle \rightarrow \left|3\right\rangle$) \cite{Moreno2011, Lopez} with fine control of $f_R$, with a 1-Hz resolution achieved through phase-locking at a signal generator (\textit{E8663B-Agilent}).

The simultaneous presence of the diode and Ti:sapphire laser beams together with quantum fluctuations are enough to initiate the FWM process \cite{Agarwal}, leading to the generation of new coherent beams at 5.23 $\mu$m (mid-infrared) and 420 nm (blue). We detect the CBL using blue filters, a pair of diffraction gratings, and a photomultiplier. As for the IR light at 5.23 $\mu$m, it is absorbed by the glass cell that contains the Rb. To amplify the efficiency of the FWM process, we use a pair of convergent lenses with focal distances of $20$ cm to increase the intensity of the two beams in the center of the Rb cell. The beams enter the cell with parallel circular polarizations \cite{Akulshin2009}. Additional details regarding the experimental setup can be seen in \cite{Moreno2019}.

\begin{figure}[htbp]
	\centering
		\includegraphics[width=0.45\textwidth]{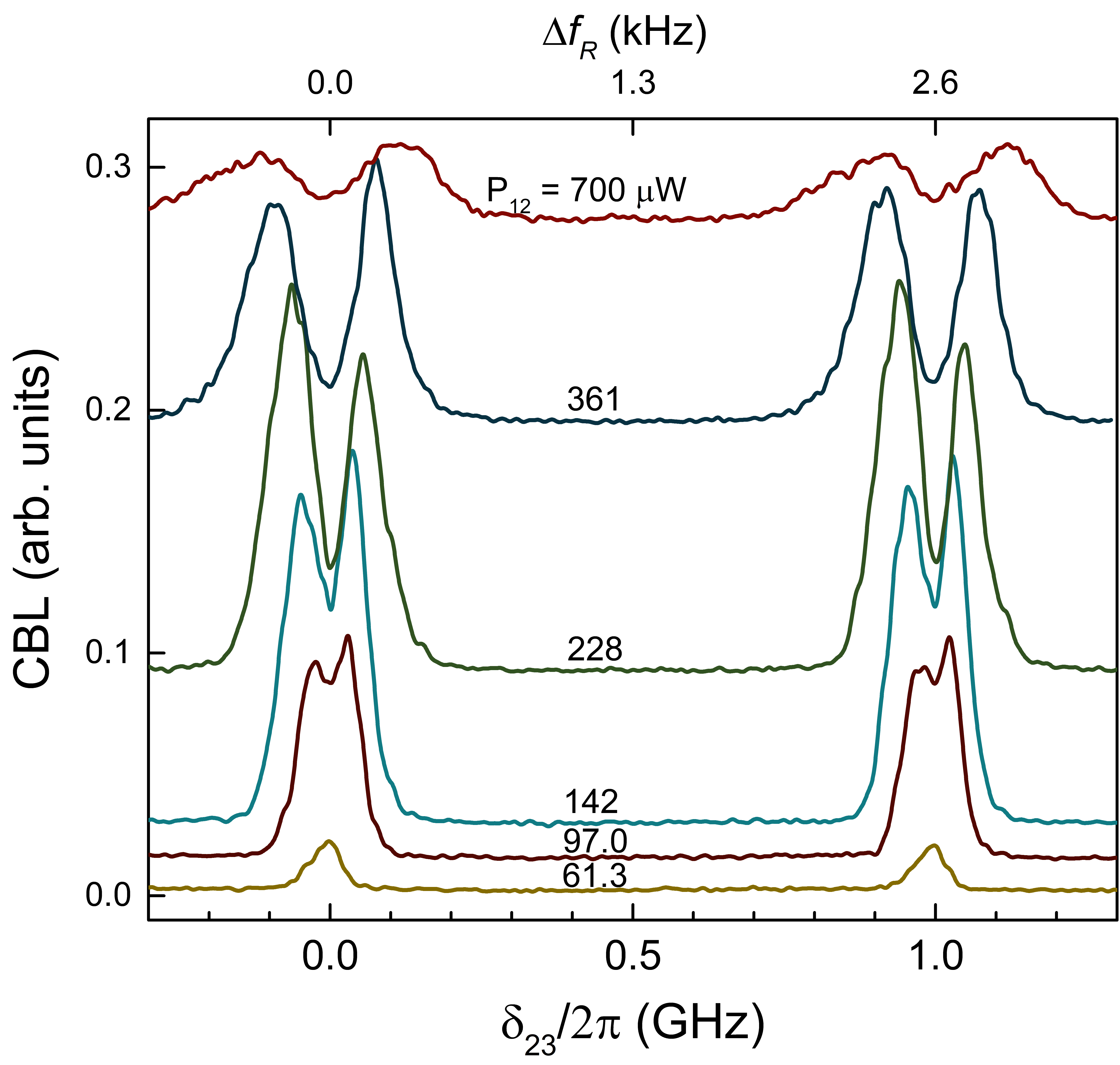}
	\caption{CBL as a function of the optical mode detuning in the $\left|2\right\rangle \rightarrow \left|3\right\rangle$ transition, for diode laser powers from 61.3 $\mu W$ up to 700  $\mu W$ (estimate at the middle of the cell). Bottom axis: detuning of the optical mode nearest of resonance. Top axis: the correspondent repetition rate variation from $f_R = 990.410$ MHz.}
	\label{fig9}
\end{figure}

In Fig. 9, we show the intensity of the CBL as a function of the Ti:sapphire optical mode detuning ($\delta_{23}$), for different powers of the diode laser. These measurements were performed with the Rb cell at a temperature of 74 $^o$C and with the diode laser on resonance ($\delta_{12} = 0$). We estimate that the power of the diode laser at the center of the cell ranges from $P_{12} = 61.3$ to 700 $\mu$W, while the power per optical mode of the Ti:sapphire laser is $P_{23} \approx 15$ $\mu$W. We estimate the power of the beams in the focal region by measuring the input and output power of the cell and applying the Beer-Lambert law. The same percentage of absorption was used to infer the power of the Ti:sapphire optical mode. We scan the mode frequency by varying the repetition rate of the pulses, as shown in the top axis of  Fig. \ref{fig9}. The pair of doublets corresponds to the FWM signal generated by two adjacent modes of the frequency comb, with a frequency difference of $\approx$ 990.4 MHz. We use this difference to calibrate the horizontal bottom axis of the curve. As expected by our numerical results presented in Fig. \ref{fig8}(a), the splitting of the doublet is linear with the square root of the diode laser power.

\begin{figure}[htbp]
	\centering
		\includegraphics[width=0.45\textwidth]{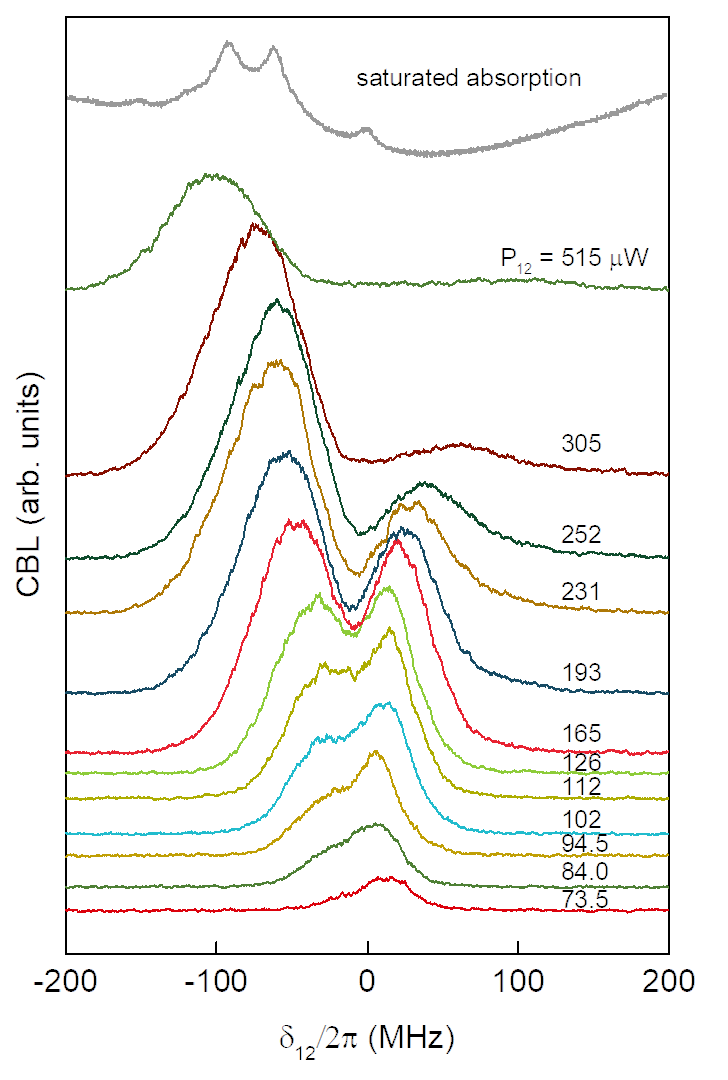}
	\caption{Normalized CBL as a function of the diode laser detuning, for a fixed Ti:sapphire laser power and diode laser powers ranging from $P_{12} = 73.5$ up to 515 $\mu$W  (estimate at the middle of the cell). The calibration of the diode laser detuning is obtained with a saturated absorption spectroscopy (top curve).}
	\label{fig10}
\end{figure}

The results for the second frequency scanning regime are in Fig. \ref{fig10}. In this case, the strong field (diode laser) sw.pdf its frequency while the weak field (one mode of the frequency comb) is locked on a frequency near resonance. There is a notable asymmetry in the doublet, with two main factors behind it: the diode laser absorption is different throughout the three hyperfine transitions of the $D_{2}$ line of Rb; the mode of the Ti:sapphire laser is not precisely on resonance. This last factor comes from the drifting off-set laser frequency, as we can only lock the repetition rate. This is in agreement with the theoretical results, as Fig. \ref{fig7}(d) indicates that even a small detuning is enough to change the symmetry of the FWM signal. As for the splitting of the doublet, we verify again a linear dependence with the square root of the diode laser power, in agreement with our numerical results of Fig. \ref{fig8}(b).

\begin{figure}[htbp]
	\centering
		\includegraphics[width=0.45\textwidth]{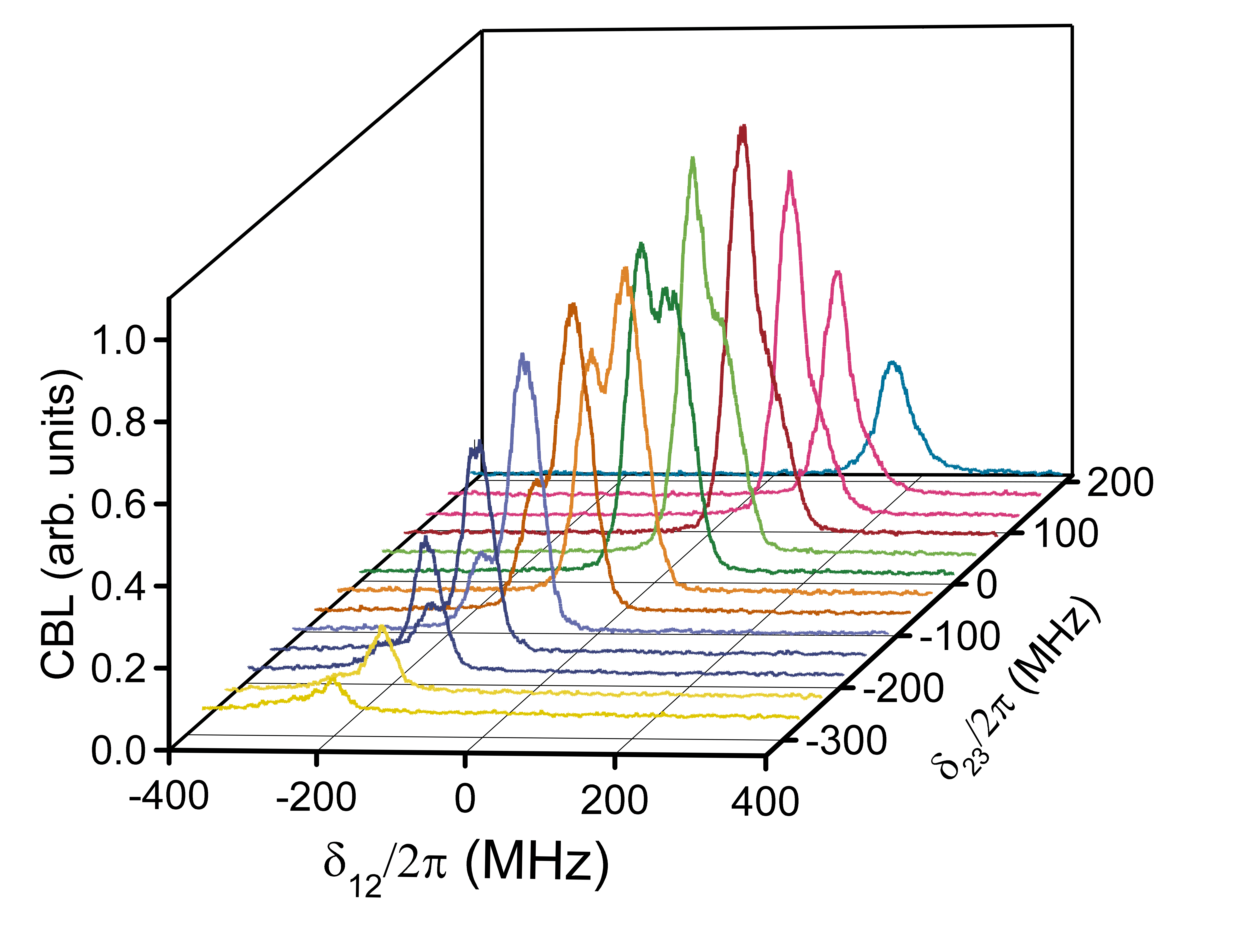}
	\caption{CBL as a function of the diode laser detuning for different repetition rates of the frequency comb. The diode laser power is $P_{12} = 112$ $\mu$W and the Ti:sapphire mode power is $P_{23} = 15$ $\mu$W (estimate at the middle of the cell).}
	\label{fig11}
\end{figure}

The graphs of Fig. \ref{fig11} confirms the theoretical results of Fig. \ref{fig7}. Whenever the lasers have the same detuning, the resonance condition is satisfied and, thus, there is a signal ($\delta_{12}-\delta_{23}=0$). Once again, this can only happen due to the inhomogeneously broadened nature of the atomic vapor, namely, there is usually a velocity group that can interact with both lasers. Moreover, if the laser in the lower transition is strong enough (in this case, the diode laser), and its frequency is varying, the AC Stark effect in combination with the Doppler profile will result in a doublet structure. The symmetry of this structure is strongly affected by the frequency position of the fixed frequency laser, as Figs. \ref{fig7}(c) and (d) predicted, and Fig. \ref{fig11} confirms.

\begin{figure}[htbp]
	\centering
		\includegraphics[width=0.49\textwidth]{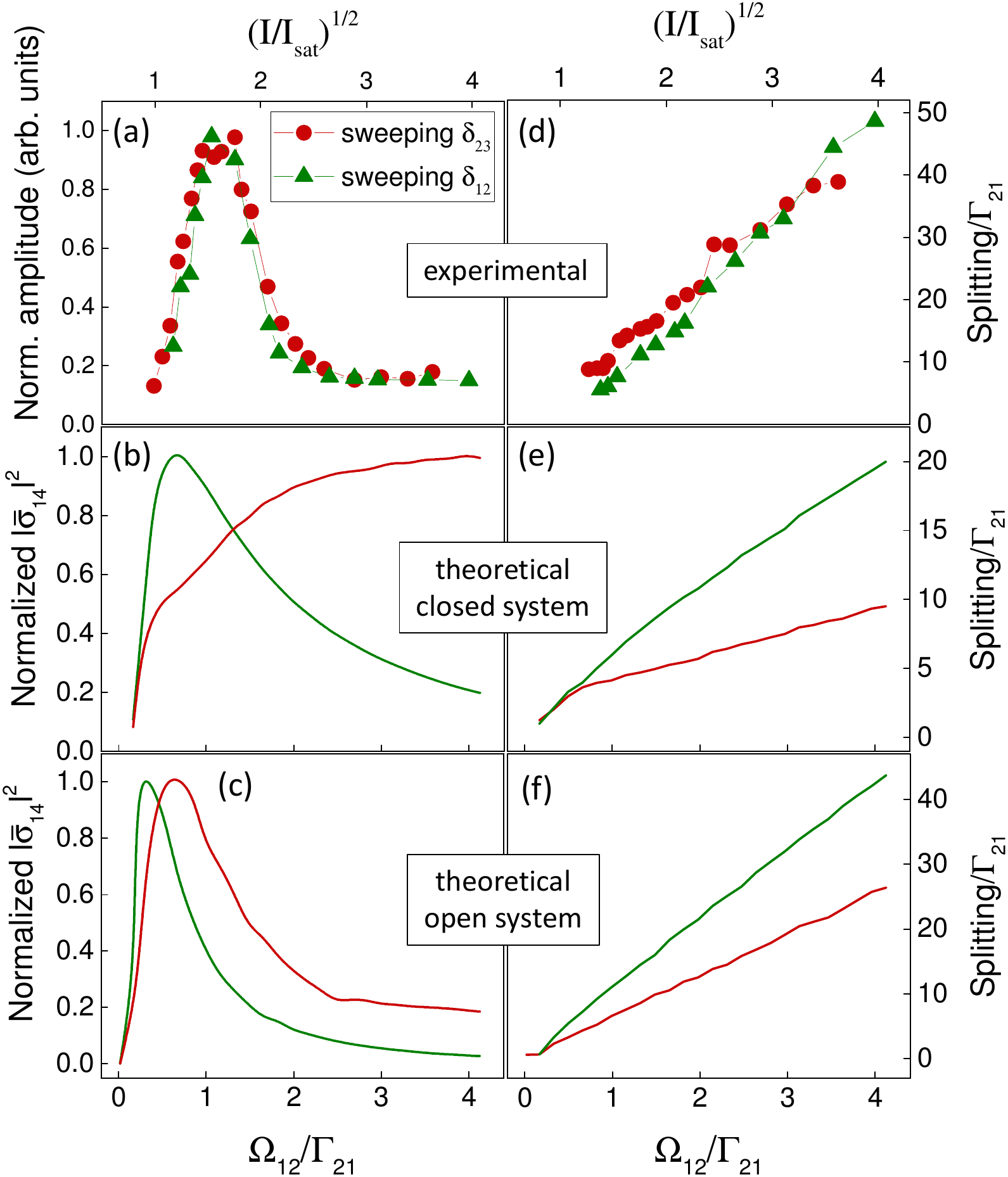}
	\caption{Experimental and theoretical curves for amplitude and splitting of the doublet as a function of square root intensity or Rabi frequency of the strong field. Red circles/lines: weak beam scanning ($\delta_{23}$); Green triangles/lines: strong beam scanning ($\delta_{12}$). (a) and (d) are the experimental results. (b) and (e) are the theoretical results for a closed system while (c) and (f), are the same results for an open system.}
	\label{fig12}
\end{figure}

In the previous section, we discussed the peak amplitude of the FWM signal for the two frequency scanning regimes. In Fig. \ref{fig12}(a), we present the experimental results for this parameter as a function of the square root of the diode laser intensity. As we can see, there is an amplitude decay for high-intensity beams, for both frequency scanning regimes. However, our theoretical model (Fig.  \ref{fig12}(b)) cannot predict the experimental behavior if the weak beam is scanning. We discussed in sec. IV that for a closed system, there will always be atoms with the proper velocity to satisfy the two-photon condition. This intensity saturation behavior of the FWM signal was observed in a pure four-level system \cite{Whiting}. However, in the experiment described here, there are three possible hyperfine transitions to the diode laser to induce. In the weak beam frequency scanning regime, the diode laser is fixed on a cyclic transition, and therefore, the system is closed. But, if the diode laser is strong enough, it will pump atoms to the open transitions, meaning that the system will no longer be closed as the atoms fall into a different fundamental hyperfine level of rubidium. To add this possibility in the model in a simple manner, we introduce a 1 MHz decay rate in the population $\rho_{22}$, allowing the loss of about 1/6 of the atoms when solving the Bloch equations. This results in the peak amplitude as a function of the strong beam Rabi frequency of Fig. \ref{fig12}(c). This way, the model achieves a behavior compatible with the experiment and reveals that the observed decay of the signal in the two frequency scanning configurations has different mechanisms behind it for each situation: optical pumping for other hyperfine levels and far detuning from resonance.

Another feature we approached in the previous section was the frequency separation between the peaks of the signal, or the ``splitting'' of the doublet. The experimental splitting seems to be the same for both frequency scanning setups, as Fig. \ref{fig12}(d) shows. However, our theoretical model not only predicts a difference between the splitting in the two scanning regimes but also gives lower values for the splitting, as Fig. \ref{fig12}(e) shows. Once again, if we consider that the system is open due to the high intensity of the strong field, these results do improve, as Fig. \ref{fig12}(f) shows. A final consideration to this is the error bar of the experimental frequency measurement. Each scanning regime uses a different, and therefore, more or less precise, calibration parameter. If the diode laser is scanning, we use the saturation absorption curve while for the Ti:sapphire scanning, the repetition of the signal, due to two consecutive frequency modes, gives the time-frequency conversion factor. So there can be a difference between the experimental splittings but masked by a systematic error in the time-frequency conversion.

\section{Conclusions}
\label{conclusions}

We have analyzed the AT splitting pattern in a dressed cascade three-level Doppler-broadening system. Our numerical calculation of the Bloch equations allows us to compare the response of the coherent blue light generated in the FWM process and the blue fluorescence given by the upper-level population for homogeneously and non-homogeneously broadened medium. In the case of homogeneous broadening, the response does not depend on whether the beams are co- or counter-propagating. The  AT doublet is present in both fluorescence and FWM signals if the weak beam frequency is sweeping. However, this doublet pattern is indistinguishable in the strong field frequency scanning regime.

On the other hand, for a Doppler-broadening medium, we need to account for the contribution of all velocity groups within the Doppler profile, resulting in a different response depending on whether the beams are copropagating or counterpropagating and which beam is sweeping. In this context, a more intriguing result is revealed when the incident beams travel in the same direction: a doublet structure is observed in the FWM signal for both scanning regimes but not in the fluorescence signal. Noteworthy, we investigate the role of the physical mechanism responsible for this doublet structure and how it depends on which scanning regime is chosen. While, in weak field scanning regime, the doublet structure is directly related to the contribution of AT effect due to all velocity groups; in the other regime, the two peaks correspond to the most distant peaks of the resonance due to two AT doublets of different velocity groups.

\section*{Acknowledgments}
\label{acknowledgments}

This work was supported by CAPES (PROEX 534/2018, No. 23038.003382/2018-39), CNPq (No. 400807/2016-5) and FAPERO (No. 01.1331.00019-0004/2014, No. 01.1331.00031-00.057/2017). A. A. C. Almeida acknowledges financial support by CNPq (No. 141103/2019-1).


\end{document}